\documentclass[%
prr,
 reprint,nodate,
superscriptaddress,
 amsmath,amssymb,
 aps,
]{revtex4-2}

\pdfoutput=1

\usepackage[utf8]{inputenc}

\usepackage{xcolor}

\usepackage{graphicx}
\usepackage{dcolumn}
\usepackage{bm}

\usepackage[bordercolor=black,backgroundcolor=yellow,linecolor=black,
textsize=scriptsize,shadow]{todonotes}

\newcommand{\mysection}{\section}

\newcommand{\mC}{\mathcal{C}}

\newcommand{\mR}{\mathcal{R}}
\newcommand{\mT}{\mathcal{T}}

\newcommand{\mO}{\mathcal{O}}

\newcommand{\tcs}{\tau,\sigma}
\newcommand{\tck}{\tau,\kappa}

\newcommand{\hr}{\middle| h \right\rangle }
\newcommand{\hl}{\left\langle h \middle|}
\newcommand{\hrr}{\left| h \right\rangle }
\newcommand{\hll}{\left\langle h \right|}

 \usepackage[colorlinks = true,
            linkcolor = blue,
            urlcolor  = blue,
            citecolor = blue,
            anchorcolor = blue]{hyperref}

\begin{document}

\author{Mario Flory}
\email{mflory@th.if.uj.edu.pl}
\affiliation{Institute of Physics, Jagiellonian University, 
30-348 Krak{\'o}w, Poland}

\author{Michal P.\ Heller} \email{michal.p.heller@aei.mpg.de}
\altaffiliation{On leave of absence from: \emph{National Centre for Nuclear Research, 02-093 Warsaw, Poland}}
\affiliation{Max Planck Institute for Gravitational Physics (Albert Einstein Institute), 14476 Potsdam-Golm, Germany}

\title{Geometry of Complexity in Conformal Field Theory}

\date{\today}

\begin{abstract}
We initiate quantitative studies of complexity in (1+1)-dimensional conformal field theories with a view that they provide the simplest setting to find a gravity dual to complexity. Our work pursues a geometric understanding of complexity of conformal transformations and embeds Fubini-Study state complexity and direct counting of stress tensor insertion in the relevant circuits in a unified mathematical language. In the former case, we iteratively solve the emerging integro-differential equation for sample optimal circuits and discuss the sectional curvature of the underlying geometry. In the latter case, we recognize that optimal circuits are governed by Euler-Arnold type equations and discuss relevant results for three well-known equations of this type in the context of complexity.


\end{abstract}

\maketitle

\mysection{Introduction} One of the most perplexing recent results in quantum gravity are holographic complexity proposals. They are convincing albeit speculative relations between volumes~\cite{Susskind:2014rva,Stanford:2014jda,Couch:2016exn} or actions~\cite{Brown:2015bva,Brown:2015lvg} in anti-de Sitter spacetimes~\cite{Maldacena:1997re,Witten:1998qj,Gubser:1998bc} and hardness of preparing states in dual quantum field theories (QFTs). 

Holographic complexity proposals not only call for a better understanding and explicit derivation, but also motivate searching for other manifestations of complexity in gravity. However, advancing the broadly defined field of holographic complexity requires expanding our knowledge of complexity in QFT. The aim of the present work is to initiate \emph{quantitative} studies of physical notions of complexity in what arguably is the most universal setting of all -- conformal field theories in 1+1 spacetime dimensions (CFTs$_{1+1}$) and local conformal transformations generated by the energy-momentum tensor. 

Complexity in its native quantum computing setting concerns hardness of approximating a given unitary using circuits composed from discrete gates acting only on a limited number of qubits (circuit complexity) or approximating a desired quantum state using such circuits acting on a simple state (state complexity). 

Motivated by holographic complexity proposals, \cite{Chapman:2017rqy,Jefferson:2017sdb} initiated studies of complexity in QFTs. These works view the preparation of a unitary operator $U$ or, upon acting on a reference state $|\mR\rangle$, also (target) state preparation $ |\mT \rangle = U |\mR\rangle$, in a continuous way as a path-ordered exponential
\begin{equation}
\label{eq.circuit}
U(\tau) =  \overleftarrow{{\cal P}} e^{- i \int_{0}^{\tau}  Q(\gamma) d \gamma}
\end{equation}
with $U(\tau = 1)$ being equal to some desired unitary $U$. In this equation, the Hermitian operator $Q(\tau)  d\tau$ is a single layer of the circuit parametrized by the parameter $\tau$ that constructs $U$. The idea used in \cite{Chapman:2017rqy,Jefferson:2017sdb} to \emph{define} complexity in a QFT appeared earlier in \cite{Nielsen:2005mn1,Nielsen:2006mn2,Nielsen:2007mn3} as a way of bounding complexity of discrete circuits acting on qubits. The relevant definition assigns a cost to $Q(\tau)$ which reflects the decomposition of $Q(\tau)$ into more elementary building blocks (\emph{gates}), each with a specified cost, and minimizes the sum of the contributions from all layers of the circuit subject to appropriate boundary conditions.
Most of the studies to date were concerned with free QFTs and claimed optimality of circuits with $Q(\tau)$ being at most quadratic in underlying bosonic or fermionic operators. Such studies, despite their simplicity, could be fine-tuned to reproduce predictions of holographic complexity proposals \cite{Chapman:2017rqy,Jefferson:2017sdb,Chapman:2018hou,Chapman:2018bqj,Caceres:2019pgf,Bernamonti:2019zyy,Bernamonti:2020bcf}.

In our eyes, studies of complexity in QFT are similar to the development of entanglement entropy in the same context. The latter also arose in connection with black hole physics~\cite{Bombelli:1986rw,Srednicki:1993im,Holzhey:1994we} and later became an independent subject with a strong quantum gravity component. One of the most fruitful seeds of progress in this discipline originated from the studies of entanglement entropy in the setting of CFTs$_{1+1}$. This includes in particular the universal result for single interval entanglement entropy in the vacuum \cite{Holzhey:1994we,Calabrese:2004eu}. Drawing a parallel from the field of entanglement entropy, which includes also the matching of the results of \cite{Holzhey:1994we,Calabrese:2004eu} by the holographic entanglement entropy~\cite{Ryu:2006bv}, CFTs$_{1+1}$ should indeed be viewed as an ideal setting for accelerating our understanding of complexity in QFTs and holography.

This vision is shared by \cite{Flory:2018akz,Caputa:2018kdj,Belin:2018bpg,Flory:2019kah,Camargo:2019isp,Erdmenger:2020sup}, and the challenges encountered in these works directly motivate our approach. These include badly posed variational problems, assigning cost predominantly to trivial $U(1)$ factors in~\eqref{eq.circuit}, or finding optimal circuits only to leading order for infinitesimal transformations. In the present work, we overcome these challenges and present a unified, geometric and \emph{quantitative} view on complexity in CFTs$_{1+1}$.

The universality of CFTs$_{1+1}$ stems from their stress-energy tensor generating the Virasoro algebra. Adopting the setting of \cite{Caputa:2018kdj,Erdmenger:2020sup}, we will be concerned with unitary circuits obtained from the exponentiation of the stress-energy tensor but employ a different way of assigning a cost to the involved operations. We focus on a CFT$_{1+1}$ defined on a Lorentzian cylinder, whose circle has a unit radius and is parametrized by the coordinate $\sigma$. We will also restrict ourselves to one copy of the Virasoro algebra. This means we will assign complexities to unitary circuits of the form~\eqref{eq.circuit} on a representation of the group of diffeomorphisms on the circle, where an operator $U(\tau)$ corresponds to a group element $f(\tcs)$ that maps the circle to itself. The function $f(\tau, \sigma)$ hence represents a sequence of diffeomorphisms of $\sigma$ interpolating between the identity $f(0, \sigma) = \sigma$ and a desired one $f(1,\sigma) \equiv f(\sigma)$. Note that we will ignore terms stemming from the central extension of the group, as these would lead to additional complex phase-factors in \eqref{eq.circuit}, which are generally considered to be irrelevant for physical notions of complexity (a problem faced in \cite{Caputa:2018kdj,Erdmenger:2020sup}). An infinitesimal layer of the circuit is generated by
\begin{equation}
\label{eq.Qcft}
Q(\tau) = \int_{0}^{2 \pi} \frac{ d\sigma}{2 \pi} \, T(\sigma)\epsilon(\tau,\sigma)\,,
\end{equation}
where $T(\sigma)$ is the right-moving component of the stress-energy tensor operator and $\epsilon(\tcs)$ is an element of the Lie-algebra defined via
 \begin{align}
 \epsilon(\tau,f(\tcs))=\dot{f}(\tcs).
 \label{epsilonf}
 \end{align}

In this letter we discuss two viable instances of cost functions. The first emerges by taking an energy eigenstate $|h \rangle$ and evaluating the Hilbert space distance traversed by the circuit defined by \eqref{eq.circuit} and~\eqref{eq.Qcft}. This is the Fubini-Study complexity defined in \cite{Chapman:2017rqy}. The second instance realizes the approach of \cite{Jefferson:2017sdb} and arises from treating $T(\sigma)$ as a one-parameter set of elementary contributions to each circuit layer and minimizing the $L_{2}$-norms of
$\epsilon(\tcs)$, $\epsilon'(\tcs)$, or a combination of both, averaged over all circuit layers.

Our approach is the first study of complexity in a generic (including large central charge $c$) CFT$_{1+1}$ that 1) does not assign cost to trivial factors (see \cite{Erdmenger:2020sup}), 2) has a well-posed variational problem for determining a transformation between two arbitrary unitaries generated by the insertions of $T(\sigma)$, 3) sheds light on the underlying geometry of circuits by probing its sectional curvatures and 4) selects the physically desirable notions of complexity based on the properties of optimal circuits. While we intend to present our main results in a concise and self-contained manner, some further discussions can be found in our companion work~\cite{Flory:2020dja}.

\mysection{Cost functions and complexity} The Fubini-Study metric a.k.a.~fidelity susceptibility arises from considering an overlap between two nearby states in the Hilbert space \cite{geomquantstates}. It is attractive from the point of view of holography and the largely open problem of physical interpretation of holographic complexity proposals, since it is known how the overlap between at least certain states in holographic QFTs manifest itself on the gravity side~\cite{Belin:2018bpg}. For a family of states $|\psi(\tau)\rangle$ parametrized by $\tau$, we can define
\begin{align}
|\langle \psi(\tau)|\psi(\tau+d\tau)\rangle|\approx 1-G_{\tau\tau}(\tau)d\tau^2+\mO(d\tau^3)
\label{fsdefinition}
\end{align}
where $G_{\tau\tau} \geq 0$ is the Fubini-Study-metric. Assume $|\psi(\tau)\rangle$ is a path on the space of states parametrized by unitary operators acting on an initial state $\hrr$,
\begin{align}
|\psi(\tau)\rangle\equiv U(\tau) \hrr.
\end{align}
The Fubini-Study metric~$G_{\tau\tau}$ becomes then the variance of $Q(\tau)$ evaluated in the state $|\psi(\tau)\rangle$ \cite{Caputa:2018kdj} or, introducing
\begin{align}
\tilde{Q}(\tau) = U(\tau)^{\dagger} Q(\tau) U(\tau),
\label{tildeQ}
\end{align}
equivalently the variance of $\tilde{Q}(\tau)$ evaluated in the state $\hrr$: $\hll\tilde{Q}^2(\tau)\hrr=\left\langle\psi(\tau)|Q^2(\tau)|\psi(\tau)\right\rangle$ by definition, and similarly for the one-point function. Note that the applications of the operators $U(\tau)$ in \eqref{tildeQ}, using  \eqref{eq.Qcft}, essentially causes a conformal transformation of the stress-energy tensor. Using the well-known transformation law and ignoring the Schwarzian term leading to an irrelevant for us phase factor, we can write \cite{Caputa:2018kdj}
\begin{equation}
\tilde{Q}(\tau) = \int_{0}^{2 \pi} \frac{ d\sigma}{2 \pi} \, T(\sigma)\frac{\dot{f}(\tcs)}{f'(\tcs)}.
\label{tildeQ2}
\end{equation}
Each trajectory through state space $|\psi(\tau)\rangle$ parametrized by $\tau\in [0,1]$ can now be assigned the total cost $\mathrm{L}_{\mathrm{FS}}$
\begin{align}
\mathrm{L}_{\mathrm{FS}}=\int_0^1d\tau \sqrt{G_{\tau\tau}(\tau)}
\label{complexity_abstract}
\end{align}
and complexity arises as its minimum subject to the appropriate boundary conditions~\cite{Chapman:2017rqy}. We should note here that the present discussion is completely general and concerns complexity of state $|\mT\rangle = U \hrr$ given a reference state $|\mR\rangle = \hrr$. Alternatively, one can view it as a definition of circuit complexity associated with a circuit representation of a unitary $U$ in which one decomposes $\tilde{Q}(\tau)$ into elementary transformations. This is similar to the notion of circuit complexity explored in~\cite{Camargo:2019isp}. 

One can alternatively define an a priori inequivalent notion of complexity based on a variance of $Q(\tau)$ in the state $\hrr$ (instead of $\left|\psi(\tau)\right\rangle$ as so far), which would be a more faithful realization of the approach~\cite{Jefferson:2017sdb}. What we mean by that is that the cost of one layer in the Fubini-Study metric depends not only on $\epsilon(\tau,\sigma)$ from a given layer, but also on what all previous layers do through the two-point function of $T$ in the \emph{evolved} state $|\psi(\tau)\rangle$. On the other hand, in the approach of~\cite{Jefferson:2017sdb} the cost of each layer depends only on $\epsilon(\tau,\sigma)$ as the insertions of $T$ would be assigned the same weight at each layer.

The discussion so far was completely general, but now we specialize to circuits defined by~\eqref{tildeQ2}. Our choice for $\hrr$, as in~\cite{Caputa:2018kdj,Erdmenger:2020sup}, is that of an energy eigenstate in the CFT$_{1+1}$ corresponding, via the operator-state correspondence, to a (quasi-)primary of the chiral dimension~$h$. To evaluate \eqref{complexity_abstract} explicitly, we follow \cite{Caputa:2018kdj,Erdmenger:2020sup} and write the variance of $\tilde{Q}$ as a bi-local integral
\begin{align}
\mathrm{L}=\int_0^1 \frac{d\tau}{2\pi} \sqrt{\iint_0^{2\pi}d\sigma 
d\kappa  \frac{\dot{f}(\tcs)}{f'(\tcs)}\frac{\dot{f}(\tck)}{f'(\tck)}  
\Pi(\sigma-\kappa)
 },
\label{complexity}
\end{align}
where $\Pi$ corresponds to a connected correlator of the stress-energy tensor in the state $\hrr$~\cite{Datta:2019jeo}
\begin{align}
&\Pi(\sigma - \kappa) =\hl T(\sigma)T(\kappa)\hr-\hl T(\sigma)\hr \hl T(\kappa)\hr
\nonumber
\\
&=\frac{c}{32 \, \sin^4{[(\sigma-\kappa)/2]}} - \frac{h}{2 \,  \sin^2{[(\sigma-\kappa)/2]}}.
\label{eq.PiTT}
\end{align}
As usual when studying geodesic motion, by requiring affine parametriztion we can move from a \textit{length-functional} \eqref{complexity_abstract} to an \textit{energy-functional} where, essentially, the square-root in \eqref{complexity_abstract} and \eqref{complexity} is removed. To do so, we note that \eqref{complexity} clearly corresponds to a geodesic problem in infinite dimensions, where summation over indices has been replaced by integration over variables $\sigma,\kappa$, $f(\tcs)$ has taken over the role of the coordinate $X^\sigma(\tau)$, and the expression $\frac{\Pi(\sigma-\kappa)}{f'(\tcs)f'(\tck)}$ plays the role of the metric $g_{\sigma\kappa}(X(\tau))$.  
The partial integro-differential equation (IDE) of motion for the circuit $f(\tcs)$ extremizing \eqref{complexity} then reads
\begin{widetext}
\begin{align}
 \int_0^{2\pi}d\sigma\Bigg(-\Pi(\sigma-\kappa)\frac{d}{d\tau}\left(\frac{\dot{f}(\tcs)}{f'(\tcs)f'(\tck)}\right)+\frac{\dot{f}(\tcs)}{f'(\tcs)}\partial_\kappa\left(\Pi(\sigma-\kappa)\frac{\dot{f}(\tck)}{f'(\tck)^2}\right)\Bigg)
 \equiv 0.
 \label{EOMf2}
\end{align}
\end{widetext}
As expected, this equation is of second order in derivatives with respect to $\tau$. This is adequate for a boundary value problem in which we envision being given an initial and final condition, $f(0,\sigma)$ and $f(1,\sigma)$, and finding the shortest circuit $f(\tcs)$ connecting these two maps. This is a notable contrast to the geometric actions studied in \cite{Caputa:2018kdj,Erdmenger:2020sup}, which lead to equations first order in $\tau$, in which generally only one initial value $f(0,\sigma)$ needs to be provided to fix a solution.  

It is the most natural for us to equate the integration kernel $\Pi$ with the connected stress-energy two-point function in the state $\hrr$~\eqref{eq.PiTT}, however we have kept our discussion more generic for a reason.
Broadly speaking, our goal in this paper is to define geodesic motion on the Virasoro-group, and this has of course already been accomplished in the framework of Euler-Arnold-type partial differential equations (PDEs) such as the Korteweg-de Vries (KdV), Camassa-Holm (CH), or Hunter-Saxton (HS) equations \cite{khesin2008geometry}. They were already discussed in the context of QFT complexity in~\cite{Caputa:2018kdj,Erdmenger:2020sup}, see also \cite{Balasubramanian:2019wgd}.
In order to show how our functional \eqref{complexity} and equation \eqref{EOMf2} fit into this framework, we note that according to \eqref{complexity}, the distance between the identity map $f=\sigma$ and a map $f=f_1(\sigma)$ is identical to the distance between $f=\sigma$ and the inverse map $F_1(\sigma)=f_1^{-1}(\sigma)$. This is easy to show by using invariance under a change of affine parameter $\tau\rightarrow s=1-\tau$ and invariance under conformal transformations $f(\tcs)\rightarrow G(f(\tcs))$. Hence, replacing the circuit $f(\tcs)$ by the inverse circuit $F(\tcs)$ in \eqref{complexity} and using the identity \cite{Caputa:2018kdj}
 \begin{align}
 \epsilon(\tau,\sigma)=-\frac{\dot{F}(\tcs)}{F'(\tcs)},
 \label{epsilon}
 \end{align}
we can write the inner product in \eqref{complexity} entirely in terms of the kernel $\Pi$ and the Lie-algebra-element $\epsilon$. Note that if $A=B\cdot C$, then $A^{-1}=C^{-1}\cdot B^{-1}$, hence replacing $f(\tcs)$ by $F(\tcs)$ in \eqref{complexity} corresponds to switching from a left- to a right-invariant metric on the Lie-group. This would yield the alternative complexity definition based on the variance of $Q$ (not $\tilde{Q}$) in the state $\hrr$ discussed below  \eqref{complexity_abstract}. 

Now, while our derivation above would suggest $\Pi$ to be the stress-energy two-point function in the state $\hrr$ \cite{Datta:2019jeo}, another choice of
\begin{align}
\Pi(\sigma-\kappa)=a \, \delta(\sigma-\kappa)+b \, \delta''(\sigma-\kappa),
\label{ultralocal}
\end{align}
with Dirac's delta function $\delta(\sigma - \kappa)$, would allow us to obtain the CH equation ($a=b=1$), the HS equation ($a=0, b=1$), and the KdV equation ($a=1,b=0$), ignoring as before the term coming from the central extension. 
Through the lens of~\cite{Jefferson:2017sdb}, \eqref{ultralocal} can be seen as assigning directly a spatially uniform cost to individual insertions of $T(\sigma)$ via $Q(\tau)$ defined in~\eqref{eq.Qcft}. In a sense, our complexity functional \eqref{complexity} corresponds to a generalization of the inner products that led to these well studied PDEs. Likewise, while these PDEs provide valid choices for a definition of complexity for the Virasoro group, they may be also regarded as simpler models for the physics encoded in the optimization problem behind the Fubini-Study complexity defined by~\eqref{EOMf2} with~\eqref{eq.PiTT}.

In the Fubini-Study case, in order to assign a well-defined finite value to \eqref{complexity} despite the pole of $\Pi$, we can make use of the method of \textit{differential regularisation}~\cite{FREEDMAN1992353,Latorre:1993xh}. This means we will (implicitly) phrase the divergent terms of the two-point-function in \eqref{complexity} as derivatives of more mildly-divergent terms and then carry the derivatives over onto the test-function $\frac{\dot{f}(\tcs)}{f'(\tcs)}\frac{\dot{f}(\tck)}{f'(\tck)} $ via integration by parts. The immediate physical consequence of this is that the metric will be degenerate. For example, if $\frac{\dot{f}(\tcs)}{f'(\tcs)}=\mathrm{const}$, the integrals in \eqref{complexity} will vanish when applying derivatives to this term due to differential regularisation.
Although the cause of some technical problems, this degeneracy of the metric is a desirable feature as it makes sure that transformations which only change the reference state by a complex phase will be assigned zero cost in terms of complexity. For this reason, we believe that the HS equation, $a=0$ in \eqref{ultralocal}, will be a more realistic model of CFT-complexity than the KdV equation studied in \cite{Caputa:2018kdj,Erdmenger:2020sup}. We discuss these issues in more detail in \cite{Flory:2020dja}.

\mysection{Optimal circuits for Fubini-Study}
The exact solutions and properties of the KdV, CH and HS equations are well studied \cite{khesin2008geometry}, and hence we will now focus again on the concrete equations of motion \eqref{EOMf2} with $\Pi$ given by~\eqref{eq.PiTT}. For this, apart from trivial circuits such as $f(\tcs)=f(\tau+\sigma)$, we do not know exact solutions. However, for boundary conditions of the form $f(0,\sigma)=\sigma$, $f(1,\sigma)=\sigma+\frac{\varepsilon}{m}\sin(m\sigma)$ with $m \in\mathbb{N}$, $\varepsilon\ll1$, we iteratively construct a circuit $f(\tcs)$ satisfying \eqref{EOMf2} order by order in $\varepsilon$. For example, for $m=1$ this yields
\begin{align}
f(\tcs) = &\sigma +\varepsilon\tau  \sin (\sigma )
\label{sinsolution}
\\
&+\varepsilon^2 \frac{c \tau ^2-c \tau +20 h \tau ^2-20 h \tau}{4 (c+8 h)} \sin (2 \sigma )+...
\nonumber
\end{align} 
and the square of the on-shell Lagrangian in \eqref{complexity} is
\begin{align}
&\iint d\sigma 
d\kappa\ \Pi(\sigma-\kappa)\   
\frac{\dot{f}(\tcs)}{f'(\tcs)} \frac{\dot{f}(\tck)}{f'(\tck)}
\label{epsilonseries}
\\
&=2 \pi ^2 \, \mathit{h} \,\varepsilon ^2+\frac{\pi ^2\, \left(3 \,\mathit{c}^2+56\, \mathit{c}\, \mathit{h}+112 \,\mathit{h}^2\right)}{96\, (\mathit{c}+8 \,\mathit{h})} \varepsilon ^4+...
\nonumber
\end{align}
We comment more on this result and how it was obtained in the appendix. Figure \ref{fig::circuit} shows a graphical representation of the circuit \eqref{sinsolution}.

\begin{figure}[t]
\includegraphics[width=0.5\textwidth]{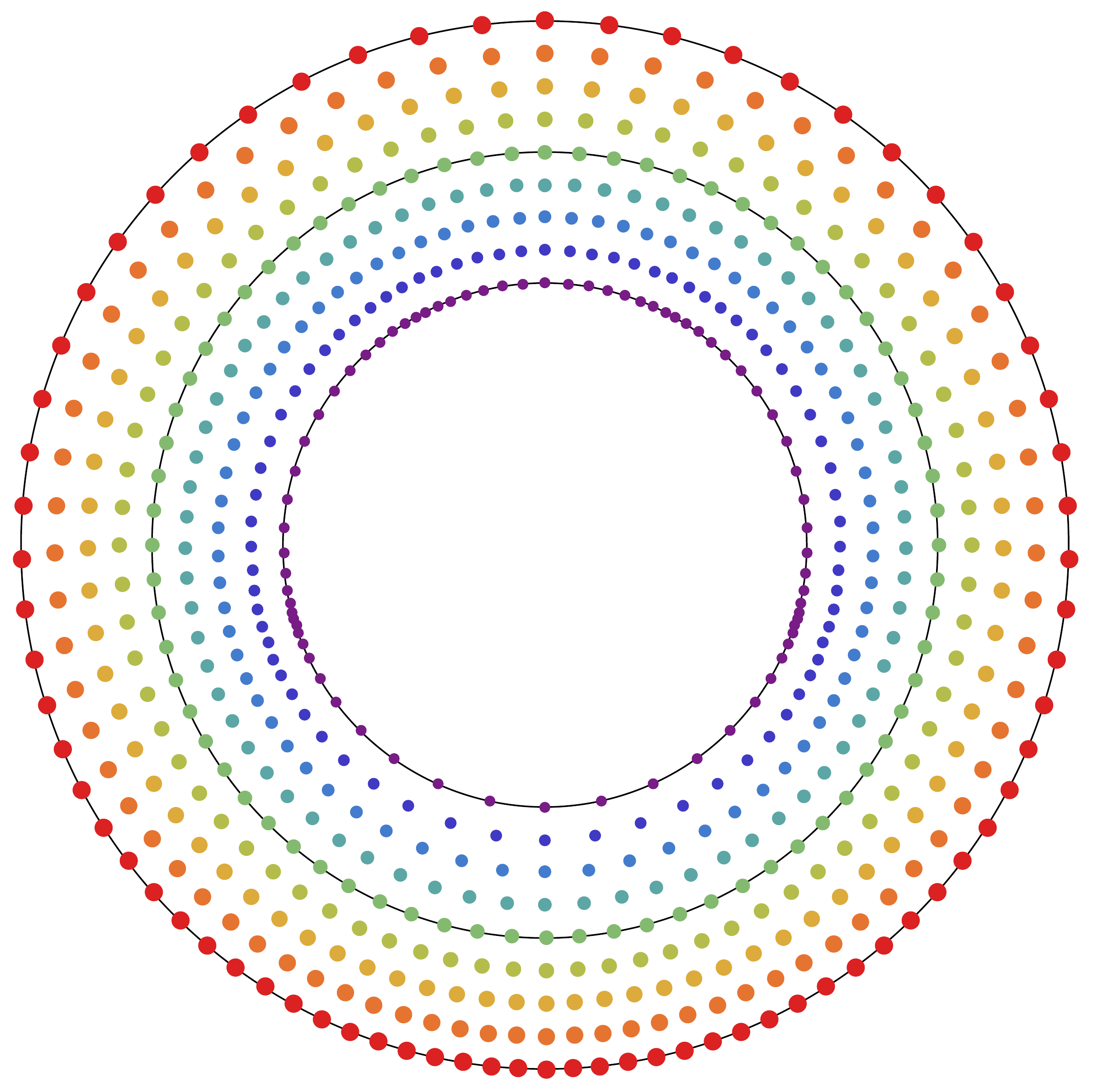}
	\caption{A graphical representation of the circuit \eqref{sinsolution}, where for simplicity we have set $c=0, h=1$ and $\epsilon=1$ in order to generate a result that is visible to the eye. For each fixed value of $\tau$, $f(\tcs)$ is a map of the circle to itself, depicted by the colored points unevenly spread out along the circles according to the function $f(\tcs)$ with evenly spaced input values for $\sigma$. The radial ordering (also emphasized by colour) corresponds to the progression of the parameter $\tau$ from $-1$ to $1$ in steps of $\Delta\tau=1/4$. The solid circles mark $\tau=-1,0,1$, respectively. At $\tau=0$, $f(\tcs)$ is the identity map by construction. "Wave breaking" happens when infinitesimally close points are mapped to the same location because $f'(\tcs)=0$. }
	\label{fig::circuit}
\end{figure}

\mysection{Geometry of complexity} In order to gain a qualitative understanding of the geometry of the Fubini-Study complexity and compare with other notions, we will instead proceed to calculate the sectional curvatures $K(u,v)$ at the identity map $f(\sigma)=\sigma$ for tangent-vectors of the form $u=sin(m\sigma)$, $v=\sin(n\sigma)$, $m,n \in\mathbb{N}$ and $m\neq n$. 
Following the analogy with finite dimensional geodesic motion, first and second derivatives of the metric can then be defined as functional derivatives, e.g. $\frac{\partial g_{\sigma\kappa}}{\partial x^\eta}\rightarrow \frac{\delta}{\delta f(\eta)}\frac{\Pi(\sigma-\kappa)}{f'(\sigma)f'(\kappa)}$
$=-\Pi(\sigma-\kappa)\left(\frac{\delta'(\sigma-\eta)}{f'(\sigma)^2f'(\kappa)}+\frac{\delta'(\kappa-\eta)}{f'(\sigma)f'(\kappa)^2}\right)$.
In order to calculate the sectional curvatures,
\begin{align}
K(u,v)=\frac{R_{\sigma\kappa\eta\omega}u^\sigma u^\eta v^\kappa v^\omega}{(u \cdot u)(v \cdot v)-(u\cdot v)^2},
\label{Kuv}
\end{align}
we would still need an analogue of the inverse metric which is needed in the definition of the Riemann-tensor $R_{\sigma\kappa\eta\omega}$. Note that in~\eqref{Kuv} the scalar product is taken using the metric $g_{\sigma\kappa}$ and we assume Einstein's summation formula involving integration. The inverse metric in question does not exist because our metric is degenerate. However, for circuits of the type~\eqref{sinsolution} which satisfy the condition $\int d\sigma \frac{\dot{f}(\tcs)}{f'(\tcs)} =0$ for all $\tau$, it is possible to show that the equation of motion \eqref{EOMf2} is left invariant when adding a non-zero constant to the two-point function $\Pi$. This creates a metric which is invertible and yet has identical (i.e.~independent of the added constant) sectional curvatures to the original metric in the tangent-planes spanned by $u=sin(m\sigma),v=\sin(n\sigma)$. We find
\begin{widetext}
 \begin{align}
K(u,v)=\frac{3}{\pi ^2 (m+n)}\left(\frac{(2 m+n) (m+2 n)}{24 h+c (m+n-1) (m+n+1)}-\frac{(2 m-n) (m+n)^2}{m \left(24 h+c \left(m^2-1\right)\right)}\right)\text{ for $m>n$.} 
\label{Kuvsolved}
\end{align}
\end{widetext}
For $0\leq h/c<\frac{4}{13}$, all $K(u,v)$ in \eqref{Kuvsolved} are negative, for $h/c\rightarrow+\infty$, all $K(u,v)$ are positive, and for generic $h>0,c>0$, only a finite number of $K(u,v)$ for small $m,n$ will be positive, while the infinitely many remaining ones will be negative. Due to a theorem from \cite{MILNOR1976293}, we know that there should always be \textit{some} tangent-planes in which $K>0$, but the result above suggests that the \textit{generic} curvature felt by the geodesic curves we are investigating will be negative unless $h/c\rightarrow\infty$. This is important, as in the study of Euler-Arnold-type geodesic equations negative sectional curvatures are related to a strong dependence of geodesics on initial conditions \cite{Arnold2014c}. For models of complexity, the necessity of negative sectional curvatures has been discussed in \cite{Brown:2016wib,Brown:2017jil}.
Curiously, the metrics underlying the KdV and CH equations lead to sectional curvatures of non-definite sign \cite{10.2307/2161606,MISIOLEK1998203}, while the geometry underlying the HS equation is positively curved \cite{Lenells07}. Furthermore, it was shown in \cite{Lenells07} that the geometry underlying the HS equation maps the group-manifold to an open subset of an $L^2$-sphere, implying geodesic incompleteness: Geodesics can leave the space of invertible maps on the circle in finite affine parameter $\tau$. From the point of view of the HS equation as a wave equation, this is related to the phenomenon of wave breaking, but from  a complexity perspective this phenomenon is harder to interpret. 
It would be fascinating to develop a better understanding of the conditions on a generic $\Pi$ under which equations of the form \eqref{EOMf2} allow for such wave-breaking in finite time $\tau$. We conjecture that this should not be the case for physically sensible choices of complexity.

\mysection{Summary and outlook} We addressed the problem of complexity of unitary operators resulting from exponentiation of the right- (or left-) moving component of the stress-energy tensor in CFT$_{1+1}$, see \eqref{eq.circuit} and~\eqref{eq.Qcft}. There are three important features of the complexity notions we consider: 
Firstly, they lead to equations of motion second order in derivatives of $\tau$, 
which allows to search for optimal circuits connecting two elements of the Virasoro group. Secondly, 
they are based on only counting nontrivial (neglecting phase factors) elementary operations with non-negative weights.  It would be interesting to see if there are circumstances under which the neglected phase factor can be interpreted as a geometric phase~\cite{Berry:1984jv}, see e.g.~\cite{Akal:2019hxa}. Thirdly, the whole construction uses explicitly spatially local gates generated by the stress-energy operator and counts either linear combinations of it via the Fubini-Study approaches or directly the usage of local operators itself.

We started with the definition of complexity based on the Fubini-Study distance \eqref{fsdefinition}, as it seems to be easier to embed in holography and has other attractive properties. This naturally led us to the Euler-Arnold type nonlinear 
equation~\eqref{EOMf2}.
Despite the intricate form of this equation of motion in an infinitely-dimensional space, we were able to find approximate solutions interpolating between the identity and its perturbation containing a single Fourier mode, see \eqref{sinsolution}. The leading term there reproduces the result of \cite{Flory:2018akz,Belin:2018bpg} for complexity change under infinitesimal conformal transformations, which is a double integral of two test functions (related to $\dot{f}$) integrated against the stress-energy two-point function as integration kernel. 
Higher order terms in \eqref{sinsolution} and \eqref{epsilonseries} are new predictions originating from the nonlinear nature of the equation of motion. We also evaluated the sectional curvatures at the identity and found that for the Fubini-Study metric in physically relevant cases it is negative in most directions, see \eqref{Kuvsolved}.

Subsequently, we looked at another possibility of defining circuit complexity that is based on explicit counting of appearances of the stress-energy tensor. This approach can be thought of as originating from a state in which the correlation function of the stress-energy tensor is ultra-local  \eqref{ultralocal}, and it would be exciting to pursue this analogy further, perhaps along the lines of~\cite{Miyaji:2015fia}. Optimal circuits in this case would be exactly described by the KdV, CH or HS equations, which were suggested as models for complexity in \cite{Caputa:2018kdj,Erdmenger:2020sup}. However, the geometries underlying these equations may have undesirable properties from a complexity perspective, such as positive sectional curvatures or geodesic incompleteness \cite{Lenells07}.

One intriguing open problem is realizing circuits given by \eqref{eq.circuit} and~\eqref{eq.Qcft} in holography, and their relation to holographic complexity proposals~\cite{Susskind:2014rva,Stanford:2014jda,Couch:2016exn,Brown:2015bva,Brown:2015lvg}. A strong hint in this direction is the agreement noted in~\cite{Belin:2018bpg} for infinitesimal conformal transformations using the results from~\cite{Flory:2018akz}.

Another way forward is to understand the circuit~\eqref{eq.circuit} with $Q(\tau)$ given by \eqref{eq.Qcft} as being realized by a CFT$_{1+1}$ in a curved geometry with $\tau$ as the physical time. This brings a parallel with the path-integral optimization program~\cite{Caputa:2017urj,Caputa:2017yrh,Camargo:2019isp}, which was, however, predominantly presented in the context of non-unitary circuits originating from the Euclidean time evolution. Another interesting issue is the question of permissible cost functions being covariant functionals of the underlying metric \cite{Camargo:2019isp}.
Finally, it is clearly important to see if inclusion of primary operators and their descendants in circuits containing the stress-energy tensor can lead to short-cuts, see~\cite{Balasubramanian:2019wgd}.  

\nocite{Bueno:2019ajd}

\begin{acknowledgements} We would like to thank Volker Schomerus for being involved in the initial part of this collaboration. In the context of this work and its companion paper~\cite{Flory:2020dja} we particularly benefited from correspondence with Diptarka Das and discussions with Pawe\l{} Caputa on \cite{Caputa:2018kdj}. Furthermore, we gratefully acknowledge Martin Ammon, Johanna Erdmenger,  Marius Gerbershagen, Romuald Janik, and Anna-Lena Weigel for conversations and correspondence, as well as Alexandre Belin, Minyong Guo, Ro Jefferson, Javier Mag\'{a}n, Ali Naseh, Onkar Parrikar, Blagoje Oblak, G\'{a}bor S\'{a}rosi and Claire Zukowski for very useful comments on both papers. The Gravity, Quantum Fields and Information (GFQI) group at AEI is supported by the Alexander von Humboldt Foundation and the Federal Ministry for Education and Research through the Sofja Kovalevskaja Award. MF was supported by the Polish National Science Centre (NCN) grant 2017/24/C/ST2/00469. MF is also grateful to the organizers of the GQFI Workshop IV, where this work was presented for the first time, and the GFQI group at AEI for its hospitality. 
\end{acknowledgements}

\appendix

\section*{Appendix} 

 In this extra section, we will further elaborate on the iterative procedure which led us to the approximate solution \eqref{sinsolution}, see also \cite{Flory:2020dja} for further details.

 Suppose we want to calculate the geodesic circuit from $f_0(\sigma)=\sigma$ at $\tau=0$ to some target map of our choice, written as
 \begin{equation}
 \label{eq.defg}
 f_1(\sigma)=\sigma+\varepsilon \, g(\sigma)    
 \end{equation} 
 at $\tau=1$ with $\varepsilon\ll1$. To lowest order in $\varepsilon$, the geodesic circuit connecting these two maps will just be the linear interpolation between them, and for higher order corrections in $\varepsilon$ we write
\begin{align}
 \label{eq.fnF}
 f(\tcs)=\sigma+\tau\, \varepsilon \, g(\sigma)+\varepsilon^2\, f^{(2)}(\tcs)+... \,.
 \end{align}
 This ansatz alone stills lead to integro-differential equations for $f^{(n)}$ at increasing orders in $\varepsilon$. In addition,  we represent $f^{(n)}(\tcs)$ as a Fourier series
 \begin{align}
 f^{(n)}(\tcs)=b_{n,0}(\tau)&+\sum_{j\in\mathbb{N}}b_{n,j}(\tau)\cos(j\,\sigma)
 \\
 &+\sum_{j\in\mathbb{N}}a_{n,j}(\tau)\sin(j\,\sigma).
 \nonumber
 \end{align}  
 Instead of one integro-differential equation of motion for $f^{(n)}(\tcs)$, we obtain an infinite number of coupled ordinary differential equations (ODEs) for the modes $b_{n,0}(\tau)$, $b_{n,j}(\tau)$ and $a_{n,j}(\tau)$. The term $b_{n,0}(\tau)$ can be set equal to zero at every order without loss of generality for reasons explained in \cite{Flory:2020dja} . The other modes should be solved for subject to the conditions
 \begin{equation}
 \label{eq.abconditions}
 a_{n,j}(0)=a_{n,j}(1)=b_{n,j}(0)=b_{n,j}(1)=0.
 \end{equation}

For the moment, we make the additional assumption that $g(\sigma)$ in \eqref{eq.defg} is proportional to one single Fourier-mode: $g(\sigma)=\sin(m\sigma)$ with some integer $m$. The benefit of this assumption is that the on-shell (squared) Lagrangian under the ansatz \eqref{eq.fnF} can be calculated explicitly order by order in $\varepsilon$, and equations of motion can then be calculated for the modes $a_{n,j}(\tau),b_{n,j}(\tau)$ by varying this expression for the Lagrangian. For the form of $g(\sigma)$ which we have chosen, and in fact for any $g(\sigma)$ which has a Fourier series that exactly terminates after a finite mode-number $m$, it can be shown that there will be an integer number $M(m,n)$ such that the modes $a_{n,j}(\tau),b_{n,j}(\tau)$ decouple from all other modes for $j\geq M(m,n)$, obtaining trivial equations $\ddot{a}_{n,j}=0=\ddot{b}_{n,j}$, which, subject to \eqref{eq.abconditions}, simply imply $a_{n,j}=0=b_{n,j}$.

Hence, at any finite order $n$ of $\varepsilon$, we only have to solve a \textit{finite} system of coupled ordinary differential equations for the $a_{n,j}(\tau),b_{n,j}(\tau)$ with $j<M(m,n)$, which is tedious but straightforwardly doable. Higher and higher orders in $\varepsilon^n$ can then be added iteratively. For the specific choice of $g(\sigma)=\sin(\sigma)$ ($m=1$), we obtained the solution \eqref{sinsolution} in this way.  

What is now the distance between $f(0,\sigma)=\sigma$ and $f(1,\sigma)=\sigma +\varepsilon\sin (\sigma )$? As the Lagrangian is conserved by affine parametrisation and the curve is parametrised such that it reaches its destination at $\tau=1$, the square of the Fubini-Study distance \eqref{complexity_abstract} is just (up to a factor $4\pi^2$) equal to the value $\mathrm{L}_{sq}$ of the square of the Lagrangian from \eqref{complexity}, i.e.~\eqref{epsilonseries}. As expected, for small $\varepsilon$ the distance $\mathrm{L}_{FS}\propto\sqrt{\mathrm{L}_{sq}}$ will be linear in $\varepsilon$. 

 We will now list a few important observations about this result and the iterative method in general:

\textbullet\ While we assume $\varepsilon\ll1$, in principle we can go to arbitrarily high orders in $\varepsilon^n$. This means that even when the target map \eqref{eq.defg} is \textit{not} infinitesimally close to the identity, we can calculate the extremal circuit to arbitrary precision, modulo possible issues arising in the presence of conjugate points, which we leave for future study. Furthermore, as the extremal circuit is a minimizer of length, even finite accuracy approximations could be useful in providing upper bounds on the distance which become increasingly tight as more and more orders of $\varepsilon$ are added.  

\textbullet\ Of course, if only an approximation to the circuit with finite precision is needed, the equations of motion might be solved numerically instead. Besides some technical problems pointed out in \cite{Flory:2020dja}, this would require a shooting or relaxation method while our iterative method is by design well adapted to the situation where initial and final conditions are given for the circuit (instead of initial position and velocity), which is the most physical setup for a problem of this kind. Also, the iterative method produces analytical results (as series-expansions in $\varepsilon$), which may be inherently an advantage in some circumstances.  

\textbullet\ The first point above concerns the issue of what the convergence radius is for the series-expansion in \eqref{epsilonseries}. For $\varepsilon=1$, $f'(1,\sigma)=1 +\cos (\sigma )$ can be zero for some $\sigma=\sigma_0$, and such maps with $f'(\sigma_0)= 0$ are not proper group elements as they would not be invertible one-to-one maps of the circle to itself. 
 A very important question is whether this boundary of the space of allowed maps can be reached in finite distance, which is indeed possible in the case of the HS equation as shown in \cite{Lenells07}. Based on our physical intuition for complexity, we would like to conjecture that this is not the case. This would require that the series in \eqref{epsilonseries} has a finite convergence radius $\varepsilon<1$ and describes an analytic function with a pole at $\varepsilon=1$. We hope to study this issue in more detail in the future. 

 \textbullet\ In the above we have set $g(\sigma)=\sin(m\sigma)$ as a simple example and proof of concept. However, our iterative procedure is much more versatile. Firstly, we could straightforwardly use any function $g(\sigma)$ for which the Fourier series terminates after a finite mode number (even though in practice it adds a lot of workload).

But even more generally, consider a target map with non-terminating Fourier series
 \begin{align}
 g(\sigma)=\sum_{m\in\mathbb{Z}}c_m e^{im\sigma}
 \label{Fseries}
 \end{align}
 For this to converge to a smooth function $g\in C^{\infty}$, we need $\sum_{m}|c_m||m|^N<\infty$ for \textit{any} integer $N$.  For concreteness, we can consider the case $c_m=\hat{c}_mq^m$ with some $0<q<1$ and $\hat{c}_m=\mO(1)$ for any $m$. In this case, we can simply replace $q$ with $\varepsilon$, and do the iterative procedure as before, imposing adequate boundary conditions on all modes at any given order. Essentially, as $\lim_{m\rightarrow\infty}c_m=0$, the Fourier series \eqref{Fseries} can be interpreted as always \textit{effectively} terminating for any given finite level of accuracy (determined by the magnitude of $\varepsilon^n$).

 \textbullet\ Of course, in our eyes there is no "best" or "right" choice of $g(\sigma)$. The most important possible achievement would be a comparison with bulk results, for example along the lines of the comparison between the leading order results of \cite{Flory:2018akz} and \cite{Belin:2018bpg}, but for higher orders in $\varepsilon$. For this task a simple function like $g(\sigma)=\sin(m\sigma)$ would be sufficient for obtaining non-trivial results, and we see no reason why a more complicated function $g(\sigma)$ would give any additional benefits beyond that.

 \textbullet\ In \eqref{epsilonseries} we worked with the square of the original Lagrangian, which is also more convenient for deriving the equations of motion. However, we have so far left ambiguous whether the actual value of the complexity should be calculated from this squared Lagrangian as in \eqref{epsilonseries}, or the original Lagrangian including the square root \eqref{complexity}. The reason for this is that assuming affine parametrization, the equations of motion following from both kinds of Lagrangian are equivalent, and so are hence their solutions and the sensitivity of the geodesic problem to initial conditions (related to the  sectional curvatures \eqref{Kuvsolved}), which was a focus of our paper. It is our philosophy that the study of such geometric properties of a given complexity proposal is as important as the motivation of the employed cost function itself, in contrast to some parts of the holographic literature that seem to end their investigations at the point where they have motivated a choice of cost function, without analysing the equations of motion, their generic solutions and underlying geometry in detail. 

See \cite{Bueno:2019ajd} for a discussion of the issue of distance functionals (i.e.~Lagrangians including a square-root) versus energy functionals (i.e.~squared Lagrangians). 
 In our case, if we have in mind a comparison with bulk results for the complexity=volume proposal along the lines of  \cite{Flory:2018akz,Belin:2018bpg}, the squared Lagrangian seems more relevant as then complexity scales linearly with the central charge instead of its square-root. In fact, our leading order result in $\varepsilon$ then trivially reproduces the results for complexity change of the ground state ($h=0$) calculated in \cite{Flory:2018akz,Belin:2018bpg} for the complexity=volume proposal and conformal transformation in only one copy of the Virasoro group (which we have exclusively considered in this paper).  Extending our methodology to two copies of the Virasoro group in the most trivial way $\mC_{total}\equiv\mC_{Vir_1}+\mC_{Vir_2}$ would unfortunately fail to reproduce the general results of \cite{Flory:2018akz,Belin:2018bpg} already at leading order. However, we leave it to future studiy whether for transformations in only one copy of the Virasoro group the volume change matches with our results beyond leading order.

\bibliography{literature}

\providecommand{\href}[2]{#2} \providecommand{\beforedoihref}{}
  \providecommand{\afterdoihref}{}\begingroup\raggedright\begin{thebibliography}{10}

\bibitem{Susskind:2014rva}
L.~Susskind, {\it {Computational Complexity and Black Hole Horizons}},
  \beforedoihref\href{http://dx.doi.org/10.1002/prop.201500092}{Fortsch.
  Phys.}\afterdoihref\  {\bf 64} (2016) 24--43
  [\href{http://arXiv.org/abs/1403.5695}{{arXiv:1403.5695}}]. [Addendum:
  Fortsch.Phys. 64, 44--48 (2016)].

\bibitem{Stanford:2014jda}
D.~Stanford and L.~Susskind, {\it {Complexity and Shock Wave Geometries}},
  \beforedoihref\href{http://dx.doi.org/10.1103/PhysRevD.90.126007}{Phys. Rev.
  D}\afterdoihref\  {\bf 90} (2014), no.~12 126007
  [\href{http://arXiv.org/abs/1406.2678}{{arXiv:1406.2678}}].

\bibitem{Couch:2016exn}
J.~Couch, W.~Fischler and P.~H. Nguyen, {\it {Noether charge, black hole
  volume, and complexity}},
  \beforedoihref\href{http://dx.doi.org/10.1007/JHEP03(2017)119}{JHEP}\afterdoihref\
  {\bf 03} (2017) 119
  [\href{http://arXiv.org/abs/1610.02038}{{arXiv:1610.02038}}].

\bibitem{Brown:2015bva}
A.~R. Brown, D.~A. Roberts, L.~Susskind, B.~Swingle and Y.~Zhao, {\it
  {Holographic Complexity Equals Bulk Action?}},
  \beforedoihref\href{http://dx.doi.org/10.1103/PhysRevLett.116.191301}{Phys.
  Rev. Lett.}\afterdoihref\  {\bf 116} (2016), no.~19 191301
  [\href{http://arXiv.org/abs/1509.07876}{{arXiv:1509.07876}}].

\bibitem{Brown:2015lvg}
A.~R. Brown, D.~A. Roberts, L.~Susskind, B.~Swingle and Y.~Zhao, {\it
  {Complexity, action, and black holes}},
  \beforedoihref\href{http://dx.doi.org/10.1103/PhysRevD.93.086006}{Phys. Rev.
  D}\afterdoihref\  {\bf 93} (2016), no.~8 086006
  [\href{http://arXiv.org/abs/1512.04993}{{arXiv:1512.04993}}].

\bibitem{Maldacena:1997re}
J.~M. Maldacena, {\it {The Large N limit of superconformal field theories and
  supergravity}},
  \beforedoihref\href{http://dx.doi.org/10.1023/A:1026654312961,
  10.4310/ATMP.1998.v2.n2.a1}{Int. J. Theor. Phys.}\afterdoihref\  {\bf 38}
  (1999) 1113--1133
  [\href{http://arXiv.org/abs/hep-th/9711200}{{arXiv:hep-th/9711200}}]. [Adv.
  Theor. Math. Phys.2,231(1998)].

\bibitem{Witten:1998qj}
E.~Witten, {\it {Anti-de Sitter space and holography}},
  \beforedoihref\href{http://dx.doi.org/10.4310/ATMP.1998.v2.n2.a2}{Adv. Theor.
  Math. Phys.}\afterdoihref\  {\bf 2} (1998) 253--291
  [\href{http://arXiv.org/abs/hep-th/9802150}{{arXiv:hep-th/9802150}}].

\bibitem{Gubser:1998bc}
S.~S. Gubser, I.~R. Klebanov and A.~M. Polyakov, {\it {Gauge theory correlators
  from noncritical string theory}},
  \beforedoihref\href{http://dx.doi.org/10.1016/S0370-2693(98)00377-3}{Phys.
  Lett.}\afterdoihref\  {\bf B428} (1998) 105--114
  [\href{http://arXiv.org/abs/hep-th/9802109}{{arXiv:hep-th/9802109}}].

\bibitem{Chapman:2017rqy}
S.~Chapman, M.~P. Heller, H.~Marrochio and F.~Pastawski, {\it {Toward a
  Definition of Complexity for Quantum Field Theory States}},
  \beforedoihref\href{http://dx.doi.org/10.1103/PhysRevLett.120.121602}{Phys.
  Rev. Lett.}\afterdoihref\  {\bf 120} (2018), no.~12 121602
  [\href{http://arXiv.org/abs/1707.08582}{{arXiv:1707.08582}}].

\bibitem{Jefferson:2017sdb}
R.~Jefferson and R.~C. Myers, {\it {Circuit complexity in quantum field
  theory}},
  \beforedoihref\href{http://dx.doi.org/10.1007/JHEP10(2017)107}{JHEP}\afterdoihref\
  {\bf 10} (2017) 107
  [\href{http://arXiv.org/abs/1707.08570}{{arXiv:1707.08570}}].

\bibitem{Nielsen:2005mn1}
M.~A. Nielsen, {\it {A geometric approach to quantum circuit lower bounds}},
  \href{http://arXiv.org/abs/quant-ph/0502070}{{arXiv:quant-ph/0502070}}.

\bibitem{Nielsen:2006mn2}
M.~A. Nielsen, M.~R. Dowling, M.~Gu and A.~M. Doherty, {\it {Quantum
  Computation as Geometry}},
  \beforedoihref\href{http://dx.doi.org/10.1126/science.1121541}{Science}\afterdoihref\
  {\bf 311} (2006) 1133--1135
  [\href{http://arXiv.org/abs/quant-ph/0603161}{{arXiv:quant-ph/0603161}}].

\bibitem{Nielsen:2007mn3}
M.~A. Nielsen and M.~R. Dowling, {\it {The geometry of quantum computation}},
  \href{http://arXiv.org/abs/quant-ph/0701004}{{arXiv:quant-ph/0701004}}.

\bibitem{Chapman:2018hou}
S.~Chapman, J.~Eisert, L.~Hackl, M.~P. Heller, R.~Jefferson, H.~Marrochio and
  R.~C. Myers, {\it {Complexity and entanglement for thermofield double
  states}},
  \beforedoihref\href{http://dx.doi.org/10.21468/SciPostPhys.6.3.034}{SciPost
  Phys.}\afterdoihref\  {\bf 6} (2019), no.~3 034
  [\href{http://arXiv.org/abs/1810.05151}{{arXiv:1810.05151}}].

\bibitem{Chapman:2018bqj}
S.~Chapman, D.~Ge and G.~Policastro, {\it {Holographic Complexity for Defects
  Distinguishes Action from Volume}},
  \beforedoihref\href{http://dx.doi.org/10.1007/JHEP05(2019)049}{JHEP}\afterdoihref\
  {\bf 05} (2019) 049
  [\href{http://arXiv.org/abs/1811.12549}{{arXiv:1811.12549}}].

\bibitem{Caceres:2019pgf}
E.~Caceres, S.~Chapman, J.~D. Couch, J.~P. Hernandez, R.~C. Myers and S.-M.
  Ruan, {\it {Complexity of Mixed States in QFT and Holography}},
  \beforedoihref\href{http://dx.doi.org/10.1007/JHEP03(2020)012}{JHEP}\afterdoihref\
  {\bf 03} (2020) 012
  [\href{http://arXiv.org/abs/1909.10557}{{arXiv:1909.10557}}].

\bibitem{Bernamonti:2019zyy}
A.~Bernamonti, F.~Galli, J.~Hernandez, R.~C. Myers, S.-M. Ruan and J.~Simón,
  {\it {First Law of Holographic Complexity}},
  \beforedoihref\href{http://dx.doi.org/10.1103/PhysRevLett.123.081601}{Phys.
  Rev. Lett.}\afterdoihref\  {\bf 123} (2019), no.~8 081601
  [\href{http://arXiv.org/abs/1903.04511}{{arXiv:1903.04511}}].

\bibitem{Bernamonti:2020bcf}
A.~Bernamonti, F.~Galli, J.~Hernandez, R.~C. Myers, S.-M. Ruan and J.~Simón,
  {\it Aspects of the first law of complexity},
  \beforedoihref\href{http://dx.doi.org/10.1088/1751-8121/ab8e66}{Journal of
  Physics A: Mathematical and Theoretical}\afterdoihref\  {\bf 53} (jul, 2020)
  294002.

\bibitem{Bombelli:1986rw}
L.~Bombelli, R.~K. Koul, J.~Lee and R.~D. Sorkin, {\it {A Quantum Source of
  Entropy for Black Holes}},
  \beforedoihref\href{http://dx.doi.org/10.1103/PhysRevD.34.373}{Phys. Rev.
  D}\afterdoihref\  {\bf 34} (1986) 373--383.

\bibitem{Srednicki:1993im}
M.~Srednicki, {\it {Entropy and area}},
  \beforedoihref\href{http://dx.doi.org/10.1103/PhysRevLett.71.666}{Phys. Rev.
  Lett.}\afterdoihref\  {\bf 71} (1993) 666--669
  [\href{http://arXiv.org/abs/hep-th/9303048}{{arXiv:hep-th/9303048}}].

\bibitem{Holzhey:1994we}
C.~Holzhey, F.~Larsen and F.~Wilczek, {\it {Geometric and renormalized entropy
  in conformal field theory}},
  \beforedoihref\href{http://dx.doi.org/10.1016/0550-3213(94)90402-2}{Nucl.
  Phys. B}\afterdoihref\  {\bf 424} (1994) 443--467
  [\href{http://arXiv.org/abs/hep-th/9403108}{{arXiv:hep-th/9403108}}].

\bibitem{Calabrese:2004eu}
P.~Calabrese and J.~L. Cardy, {\it {Entanglement entropy and quantum field
  theory}},
  \beforedoihref\href{http://dx.doi.org/10.1088/1742-5468/2004/06/P06002}{J.
  Stat. Mech.}\afterdoihref\  {\bf 0406} (2004) P06002
  [\href{http://arXiv.org/abs/hep-th/0405152}{{arXiv:hep-th/0405152}}].

\bibitem{Ryu:2006bv}
S.~Ryu and T.~Takayanagi, {\it {Holographic derivation of entanglement entropy
  from AdS/CFT}},
  \beforedoihref\href{http://dx.doi.org/10.1103/PhysRevLett.96.181602}{Phys.
  Rev. Lett.}\afterdoihref\  {\bf 96} (2006) 181602
  [\href{http://arXiv.org/abs/hep-th/0603001}{{arXiv:hep-th/0603001}}].

\bibitem{Flory:2018akz}
M.~Flory and N.~Miekley, {\it {Complexity change under conformal
  transformations in AdS$_{3}$/CFT$_{2}$}},
  \beforedoihref\href{http://dx.doi.org/10.1007/JHEP05(2019)003}{JHEP}\afterdoihref\
  {\bf 05} (2019) 003
  [\href{http://arXiv.org/abs/1806.08376}{{arXiv:1806.08376}}].

\bibitem{Caputa:2018kdj}
P.~Caputa and J.~M. Magan, {\it {Quantum Computation as Gravity}},
  \beforedoihref\href{http://dx.doi.org/10.1103/PhysRevLett.122.231302}{Phys.
  Rev. Lett.}\afterdoihref\  {\bf 122} (2019), no.~23 231302
  [\href{http://arXiv.org/abs/1807.04422}{{arXiv:1807.04422}}].

\bibitem{Belin:2018bpg}
A.~Belin, A.~Lewkowycz and G.~Sárosi, {\it {Complexity and the bulk volume, a
  new York time story}},
  \beforedoihref\href{http://dx.doi.org/10.1007/JHEP03(2019)044}{JHEP}\afterdoihref\
  {\bf 03} (2019) 044
  [\href{http://arXiv.org/abs/1811.03097}{{arXiv:1811.03097}}].

\bibitem{Flory:2019kah}
M.~Flory, {\it {WdW-patches in AdS$_{3}$ and complexity change under conformal
  transformations II}},
  \beforedoihref\href{http://dx.doi.org/10.1007/JHEP05(2019)086}{JHEP}\afterdoihref\
  {\bf 05} (2019) 086
  [\href{http://arXiv.org/abs/1902.06499}{{arXiv:1902.06499}}].

\bibitem{Camargo:2019isp}
H.~A. Camargo, M.~P. Heller, R.~Jefferson and J.~Knaute, {\it {Path integral
  optimization as circuit complexity}},
  \beforedoihref\href{http://dx.doi.org/10.1103/PhysRevLett.123.011601}{Phys.
  Rev. Lett.}\afterdoihref\  {\bf 123} (2019), no.~1 011601
  [\href{http://arXiv.org/abs/1904.02713}{{arXiv:1904.02713}}].

\bibitem{Erdmenger:2020sup}
J.~Erdmenger, M.~Gerbershagen and A.-L. Weigel, {\it {Complexity measures from
  geometric actions on Virasoro and Kac-Moody orbits}},
  \beforedoihref\href{http://dx.doi.org/10.1007/JHEP11(2020)003}{JHEP}\afterdoihref\
  {\bf 11} (2020) 003
  [\href{http://arXiv.org/abs/2004.03619}{{arXiv:2004.03619}}].

\bibitem{Flory:2020dja}
M.~Flory and M.~P. Heller, {\it {Conformal field theory complexity from
  Euler-Arnold equations}},
  \href{http://arXiv.org/abs/2007.11555}{{arXiv:2007.11555}}.

\bibitem{geomquantstates}
I.~Bengtsson and K.~Życzkowski, {\em Geometry of Quantum States: An
  Introduction to Quantum Entanglement}.
\newblock Cambridge University Press, 2~ed., 2017.

\bibitem{Datta:2019jeo}
S.~Datta, P.~Kraus and B.~Michel, {\it {Typicality and thermality in 2d CFT}},
  \beforedoihref\href{http://dx.doi.org/10.1007/JHEP07(2019)143}{JHEP}\afterdoihref\
  {\bf 07} (2019) 143
  [\href{http://arXiv.org/abs/1904.00668}{{arXiv:1904.00668}}].

\bibitem{khesin2008geometry}
B.~Khesin and R.~Wendt, {\em The Geometry of Infinite-Dimensional Groups}.
\newblock A series of modern surveys in mathematics. Springer Berlin
  Heidelberg, 2008.

\bibitem{Balasubramanian:2019wgd}
V.~Balasubramanian, M.~Decross, A.~Kar and O.~Parrikar, {\it {Quantum
  Complexity of Time Evolution with Chaotic Hamiltonians}},
  \beforedoihref\href{http://dx.doi.org/10.1007/JHEP01(2020)134}{JHEP}\afterdoihref\
  {\bf 01} (2020) 134
  [\href{http://arXiv.org/abs/1905.05765}{{arXiv:1905.05765}}].

\bibitem{FREEDMAN1992353}
D.~Z. Freedman, K.~Johnson and J.~Latorre, {\it Differential regularization and
  renormalization: a new method of calculation in quantum field theory},
  \beforedoihref\href{http://dx.doi.org/https://doi.org/10.1016/0550-3213(92)90240-C}{Nuclear
  Physics B}\afterdoihref\  {\bf 371} (1992), no.~1 353 -- 414.

\bibitem{Latorre:1993xh}
J.~I. Latorre, C.~Manuel and X.~Vilasis-Cardona, {\it {Systematic differential
  renormalization to all orders}},
  \beforedoihref\href{http://dx.doi.org/10.1006/aphy.1994.1037}{Annals
  Phys.}\afterdoihref\  {\bf 231} (1994) 149--173
  [\href{http://arXiv.org/abs/hep-th/9303044}{{arXiv:hep-th/9303044}}].

\bibitem{MILNOR1976293}
J.~Milnor, {\it Curvatures of left invariant metrics on lie groups},
  \beforedoihref\href{http://dx.doi.org/https://doi.org/10.1016/S0001-8708(76)80002-3}{Advances
  in Mathematics}\afterdoihref\  {\bf 21} (1976), no.~3 293 -- 329.

\bibitem{Arnold2014c}
V.~I. Arnold, {\em Exponential scattering of trajectories and its
  hydrodynamical applications}, pp.~419--427.
\newblock Springer Berlin Heidelberg, Berlin, Heidelberg, 2014.

\bibitem{Brown:2016wib}
A.~R. Brown, L.~Susskind and Y.~Zhao, {\it {Quantum Complexity and Negative
  Curvature}},
  \beforedoihref\href{http://dx.doi.org/10.1103/PhysRevD.95.045010}{Phys. Rev.
  D}\afterdoihref\  {\bf 95} (2017), no.~4 045010
  [\href{http://arXiv.org/abs/1608.02612}{{arXiv:1608.02612}}].

\bibitem{Brown:2017jil}
A.~R. Brown and L.~Susskind, {\it {Second law of quantum complexity}},
  \beforedoihref\href{http://dx.doi.org/10.1103/PhysRevD.97.086015}{Phys. Rev.
  D}\afterdoihref\  {\bf 97} (2018), no.~8 086015
  [\href{http://arXiv.org/abs/1701.01107}{{arXiv:1701.01107}}].

\bibitem{10.2307/2161606}
G.~Misiołek, {\it Conjugate points in the bott-virasoro group and the kdv
  equation},  Proceedings of the American Mathematical Society {\bf 125}
  (1997), no.~3 935--940.

\bibitem{MISIOLEK1998203}
G.~Misiolek, {\it A shallow water equation as a geodesic flow on the
  bott-virasoro group},
  \beforedoihref\href{http://dx.doi.org/https://doi.org/10.1016/S0393-0440(97)00010-7}{Journal
  of Geometry and Physics}\afterdoihref\  {\bf 24} (1998), no.~3 203 -- 208.

\bibitem{Lenells07}
J.~Lenells, {\it The hunter-saxton equation describes the geodesic flow on a
  sphere},
  \beforedoihref\href{http://dx.doi.org/10.1016/j.geomphys.2007.05.003}{Journal
  of Geometry and Physics}\afterdoihref\  {\bf 57} (09, 2007).

\bibitem{Berry:1984jv}
M.~V. Berry, {\it {Quantal phase factors accompanying adiabatic changes}},
  \beforedoihref\href{http://dx.doi.org/10.1098/rspa.1984.0023}{Proc. Roy. Soc.
  Lond. A}\afterdoihref\  {\bf A392} (1984) 45--57.

\bibitem{Akal:2019hxa}
I.~Akal, {\it {Reflections on Virasoro circuit complexity and Berry phase}},
  \href{http://arXiv.org/abs/1908.08514}{{arXiv:1908.08514}}.

\bibitem{Miyaji:2015fia}
M.~Miyaji, T.~Numasawa, N.~Shiba, T.~Takayanagi and K.~Watanabe, {\it
  {Continuous Multiscale Entanglement Renormalization Ansatz as Holographic
  Surface-State Correspondence}},
  \beforedoihref\href{http://dx.doi.org/10.1103/PhysRevLett.115.171602}{Phys.
  Rev. Lett.}\afterdoihref\  {\bf 115} (2015), no.~17 171602
  [\href{http://arXiv.org/abs/1506.01353}{{arXiv:1506.01353}}].

\bibitem{Caputa:2017urj}
P.~Caputa, N.~Kundu, M.~Miyaji, T.~Takayanagi and K.~Watanabe, {\it {Anti-de
  Sitter Space from Optimization of Path Integrals in Conformal Field
  Theories}},
  \beforedoihref\href{http://dx.doi.org/10.1103/PhysRevLett.119.071602}{Phys.
  Rev. Lett.}\afterdoihref\  {\bf 119} (2017), no.~7 071602
  [\href{http://arXiv.org/abs/1703.00456}{{arXiv:1703.00456}}].

\bibitem{Caputa:2017yrh}
P.~Caputa, N.~Kundu, M.~Miyaji, T.~Takayanagi and K.~Watanabe, {\it {Liouville
  Action as Path-Integral Complexity: From Continuous Tensor Networks to
  AdS/CFT}},
  \beforedoihref\href{http://dx.doi.org/10.1007/JHEP11(2017)097}{JHEP}\afterdoihref\
  {\bf 11} (2017) 097
  [\href{http://arXiv.org/abs/1706.07056}{{arXiv:1706.07056}}].

\bibitem{Bueno:2019ajd}
P.~Bueno, J.~M. Magán and C.~Shahbazi, {\it {Complexity measures in QFT and
  constrained geometric actions}},
  \href{http://arXiv.org/abs/1908.03577}{{arXiv:1908.03577}}.

\end{thebibliography}\endgroup
\bibliographystyle{bibstyl}

\end{document}